%Paper: cond-mat/9405017
%From: zynda@guinness.ias.edu (Erika Zynda)
%Date: Fri, 6 May 94 09:09:48 EDT
%Date (revised): Tue, 21 Jun 94 09:00:29 EDT

\input phyzzx
\nonstopmode
\nopubblock
\sequentialequations
\twelvepoint
\overfullrule=0pt
\tolerance=5000

\line{\hfill }
\line{\hfill PUPT 1466, IASSNS 94/25}
\line{\hfill cond-mat/9405017}
\line{\hfill May 1994}

\titlepage
\title{Exclusion Statistics: Low-Temperature Properties,
Fluctuations, Duality, Applications}

\author{Chetan Nayak\foot{Research supported in part by a Fannie
and John Hertz Foundation fellowship.~~~nayak@puhep1.princeton.edu}}
\vskip .2cm

\centerline{{\it Department of Physics}}
\centerline{{\it Joseph Henry Laboratories }}
\centerline{{\it Princeton University}}
\centerline{{\it Princeton, N.J. 08544}}

\author{Frank Wilczek\foot{Research supported in part by DOE grant
DE-FG02-90ER40542.~~~WILCZEK@SNS.IAS.EDU}}
\vskip.2cm

\centerline{{\it School of Natural Sciences}}
\centerline{{\it Institute for Advanced Study}}
\centerline{{\it Olden Lane}}
\centerline{{\it Princeton, N.J. 08540}}
\endpage

\abstract{We derive some physical
properties of ideal assemblies of
identical particles
obeying generalized exclusion statistics.  We discuss fluctuations,
and in this connection point out a fundamental contrast to conventional
quantum statistics. We demonstrate a
duality relating the distribution of
particles at statistics $g$ to the distribution of holes
at statistics $1/g$.  We suggest applications to Mott insulators.}

\endpage

\REF\haldane{F.D.M. Haldane, {\it Phys. Rev. Lett}. {\bf 67} (1991) 937}

\REF\wu{Y.S. Wu, University of Utah Preprint, March 1994.}

\REF\ouvry{S. Ouvry, {\it Phys. Rev. Lett}. {\bf 72} (1994) 600.}

\REF\bernard{D. Bernard and Y.S. Wu, University of Utah and
Saclay Preprint, April 1994.}

\REF\leinaas{J.M. Leinaas and J.Myrheim, {\it Nuovo Cim}.
{\bf 370} (1977) 1; see
also G. Goldin, R. Menikoff, and D. Sharp,
{\it J. Math. Phys.} {\bf 22} (1981)
1664; {\it Phys. Rev. Lett.} {\bf 51} (1983) 2246.}

\REF\wilczek{F. Wilczek, {\it Phys. Rev. Lett}.
{\bf 48} (1982) 1144;
{\bf 49} (1982) 957;
{\it Fractional Statistics and Anyon Superconductivity} (World Scientific,
1990).}

\REF\murthy {M.V.N. Murthy and R. Shankar, Madras preprint IMSc-94/01
(January, 1994).}

\REF\asw{D.P. Arovas, J.R. Schrieffer, F. Wilczek, and A. Zee,
{\it Nucl. Phys}. {\bf B 251}, (1985) 117.}

\REF\schrodinger{E. Schrodinger, {\it Statistical Thermodynamics\/}
(Cambridge University Press, 1964).}

\REF\csrefs{R. Stanley, Adv. Math. 77, (1989) 76;
M. Gaudin, Saclay preprint SPhT-92-158;
D. Bernard, V. Pasquier and Serban, hep-th/9404050;
J. Minahan and A. Polychronakos, hep-th/9404192.}

\REF\iso{S. Iso, D. Karabali and B. Sakita, {\it Nuclear Physics\/}
{\bf B388} (1992) 700;
{\it Phys. Lett}. {\bf B296} (1992) 143;
S. Iso, Tokyo preprint UT-676 (April, 1994).}

\REF\luttinger{See F. D. M. Haldane, {\it J. Phys}.
{\bf C14} (1981) 2585.}

\REF\slawek{A related proposal, which seems however to differ
significantly (most notably, in having a shift in the chemical
potential of first order in $T$, has been put forward by
J. Spalek and W. Wojcik, {\it Phys. Rev}. {\bf B37} (1988) 1532.}

\REF\ginz{A. Junod, in {\it Physical Properties of High Temperature
Superconductors II} ed. D. Ginsberg (World Scientific, 1990) -- see
especially page 47.}

\REF\highT{W. Putikka, R. Glenister, R. Singh, and H. Tsunetsugu,
FSU-ETH preprint NHML-93-988 ETH-TH 9333, September 1993.}

\REF\var{C. Gros and R. Valenti, Dortmund preprint (April, 1994).}

\chapter{Exclusion Statistics}

Several years ago, Haldane [\haldane] introduced the notion of
generalized or fractional exclusion statistics,
interpolating between bosonic and fermionic statistics.
Motivated by the properties of quasiparticles
in the fractional quantum Hall effect
and in one dimensional inverse-square exchange
spin chains, he defined the statistics, $g$ of a particle
by:
$$ g = -{{{d_{N+\Delta N}}-{d_N}}\over{\Delta N}}\eqn\defg $$
where $N$ is the number of particles and ${d_N}$ is
the dimension of the one-particle Hilbert space obtained
by holding the coordinates of $N-1$ particles fixed. Since
any number of bosons can occupy a given state,
${{d_{N+\Delta N}}={d_N}}$, and hence $g=0$. By contrast, the
Pauli exclusion principle implies that $g=1$ for fermions.
Particles with intermediate statistics -- Laughlin
quasiparticles with $g=1/m$ and 1-D spinons with $g=1/2$
were the two examples given in [\haldane]
-- satisfy a generalized exclusion principle.  Recently Wu [\wu ]
has
discussed the statistical mechanics of particles obeying a generalized
exclusion principle {\it locally in phase space\/} (see below).
Henceforth we shall call such particles $g$-ons.
Ouvry [\ouvry ] had previously discussed related
statistical statistical distributions in the context of
anyons in a strong magnetic field.  Bernard and Wu [\bernard ]
have shown that excitations in the Calogero-Sutherland
model obey this statistical mechanics.

An earlier form of exotic statistics --
anyons [\leinaas, \wilczek] -- has proved very
influential in the study of two-dimensional systems.
Anyons are particles whose wavefunctions aquire an arbitrary
phase $e^{i\theta}$ when two of them are braided.
Unlike fractional exclusion statistics particles,
anyons are special to two dimensions since in higher dimensions
two exchanges can be continuously deformed
to no exchange.
At first glance, exclusion statistics seem to have little
to do with the braiding properties
of particle trajectories which are the starting point for
anyons.  However in a recent important
paper, Murthy and Shankar [\murthy]
showed that anyons {\it do\/} satisfy a generalized
exclusion principle (contrary to
Haldane [\haldane ]).  They did this
by relating the
exclusion statistics parameter,
$g$, to the high-temperature
limit of the second virial coefficient, which is non-trivial
for an anyon gas [\asw ]. The linchpin of their argument
is that in a theory with a high-energy cutoff
(\eg\ any continuum model),
the transmutation of statistics by
attaching flux tubes [\wilczek ] will generally
push some states beyond
the cutoff, thereby reducing the Hilbert space dimension.
This generates
a fractional exclusion statistics that persists even
as the cutoff is taken to infinity.

In this paper, we investigate the elementary statistical
mechanics of fractional exclusion statistics particles further.
As an example, we calculate the low-temperature specific
heat of an ideal gas of ${1\over 2}$-ons both for
a conserved and a non-conserved particle number.
In the former case, we find a surprise:
the chemical potential appears
to be independent of temperature to $O(T)$.
We consider fluctuations in phase space occupancy for $g$-ons, find
that they exhibit subtleties not present for bosons or fermions, and on
this basis emphasize that one needs much stronger assumptions than
Haldane's to justify $g$-on statistical mechanics.
We also display a duality transformation
that relates statistics $g$
particles to statistics
$1/g$ holes. Finally, we attempt to make it
plausible
that electrons in Mott-Hubbard systems described by
the $t-J$ model might obey $g=2$ statistical mechanics;
by duality, the holes will
then obey $g=1/2$ statistical mechanics
with a modified effective temperature.

We now briefly recall the framework of $g$-on statistical mechanics
[\haldane ] [\wu ], to fix notations and for later use.
The statistical distribution of an ideal gas of
$g$-ons
can be obtained by maximizing the entropy subject to
the constraints of fixed particle number and energy, as in
[\schrodinger ].
We imagine dividing the one-particle states into a large
number of cells with $k\gg 1$ states in the each cell
and then count the number of configurations with
$n_i$ particles in the $i^{th}$ cell. An elementary
combinatorial argument gives:
$${e^S} = {\prod_i}\,{{{(d_{n_i}}+{n_i}-1)!}
\over{{n_i}!\,({d_{n_i}}-1)!}}\eqn\entcount$$
where ${d_{n_i}}$ is the dimension of the one-particle
Hilbert space in the $i^{th}$ cell with the coordinates of
${n_i}-1$ of the particles held constant.
{\it If\/} one can apply the definition \defg\
{\it locally in phase space\/} -- a big assumption that we
tentatively adopt as the working definition of $g$-ons, but
will need to discuss critically and refine shortly -- then
$${d_{n_i}} = k-g({n_i}-1)\eqn\opspace$$
Hence we must minimize the quantity,
$$\eqalign{{\bar S} &= {\sum_i}(k+(1-g)({n_i}-1))\ln(k+(1-g){n_i}-1))
- {n_i}\ln{n_i}\cr
&\qquad - (k-g({n_i}-1)-1)\ln(k-g({n_i}-1)-1)
- \beta{\epsilon_i}{n_i} + \beta\mu{n_i}\cr}\eqn\entbar$$
with respect to the occupation numbers, ${n_i}$;
$\beta$ and $\beta\mu$
are the Lagrange multipliers which enforce the constraints
of fixed energy and particle number.  By standard arguments
[\schrodinger ] one can show that $\beta$ is the inverse
temperature and $\mu$ the chemical potential.
Differentiating with
respect to ${n_i}$, we obtain (neglecting terms of $O(1)$
which are negligible compared to $\ln k$),
$${\partial{\bar S}\over{\partial{n_i}}} =
(1-g)\ln(k+(1-g){n_i}) - \ln{n_i} + g\ln(k-g{n_i})
- \beta({\epsilon_i} - \mu)\eqn\minim$$
Upon exponentiation this yields,
$${{\bar n}_i}\,{e^{\beta({\epsilon_i}-\mu)}} =
{\Bigl(1-g{{\bar n}_i}\Bigr)^g}
{\Bigl(1+(1-g){{\bar n}_i}\Bigr)^{1-g}}
\eqn\impldistfcn$$
where we have written ${{\bar n}_i}={{n_i}\over{k}}$ for short.
This is the fundamental equation that implicitly
defines the distribution function for $g$-ons.
In general it can only
be solved numerically, but for special cases including
$g=0,1,{1\over2},2,{1\over3},3,{1\over4},4$
it can be solved analytically.  The
$g=1/2$ distribution function is:
$${({{\bar n}_i})_{g=1/2}} =
{2\over{\Bigl(1+4e^{2\beta({\epsilon_i}-\mu)}\Bigr)^{1/2}}}
\eqn\semiondist$$

\chapter{Low-Temperature Properties}

At $T=0$
the distribution function vanishes for $\epsilon>\mu$ and
takes the value $2$ at $\epsilon<\mu$. More generally, one
can show by inspection of \impldistfcn\  that at $T=0$
$${({{\bar n}_i})_g} = \cases{{1\over g},
&if $\epsilon<\mu$;\cr 0,&if $\epsilon>\mu$.\cr}
\eqn\zerotdist$$
It is quite striking that at $T=0$ particles of general
exclusion statistics exhibit a ``Fermi'' surface.
This fact dictates the
low-temperature thermodynamics of systems of these
particles when the particle number is conserved.

We now develop an expansion in powers
of the temperature -- analogous to the Sommerfeld
expansion for fermions -- for the thermodynamic
functions of $g$-ons.

The Lagrange multipliers $\beta$ and $\mu$ are determined by
by the constraints
$$N = {\sum_i}{n_i} = V {\int}{{{d^D}p}\over(2\pi)^D}
\,{{{\bar n}_g}(\epsilon(p)-\mu)}
\eqn\numconst$$
$$E = {\sum_i}{\epsilon_i}{n_i} = V {\int}{{{d^D}p}\over(2\pi)^D}
\,{\epsilon(p)}{{{\bar n}_g}(\epsilon(p)-\mu)}
\eqn\enerconst$$
If the single particle energies are of the
form $\epsilon(p)=a{p^n}$, then these can be rewritten:
$${N\over{\alpha V}} = {\int}{d\epsilon}\,{{\epsilon}^{{D\over n}-1}}
\,{{{\bar n}_g}(\epsilon-\mu)}
\eqn\numeq$$
$${E\over{\alpha V}} = {\int}{d\epsilon}\,{{\epsilon}^{D\over n}}
\,{{{\bar n}_g}(\epsilon-\mu)}
\eqn\enereq$$
The uninteresting constant factors have been lumped into $\alpha$.
The right hand sides of these equations are both of the form
$$ {\int_0^\infty}{d\epsilon} {{\epsilon}^p}
\,{{{\bar n}_g}(\epsilon-\mu)}\eqn\intform$$
Such integrals are expanded by changing variables to
$\eta = \beta\epsilon$ and shifting the integral by $\beta\mu$,
$$\eqalign{{{\beta}^{p+1}} {\int_0^\infty}{d\epsilon} {{\epsilon}^p}
\,{{{\bar n}_g}(\epsilon-\mu)} &= {\int_{-\beta\mu}^\infty}
{d\eta}{(\eta+\beta\mu)^p} {{{\bar n}_g}(\eta)},\cr
&={\int_0^\infty}
{d\eta}{(\eta+\beta\mu)^p} {{{\bar n}_g}(\eta)} + {\int_{-\beta\mu}^0}
{d\eta}{(\eta+\beta\mu)^p} {{{\bar n}_g}(\eta)},\cr
&= {\int_0^\infty}
{d\eta}{(\eta+\beta\mu)^p} {{{\bar n}_g}(\eta)}
+ {\int_{-\beta\mu}^0} {d\eta}{(\eta+\beta\mu)^p}
 \Bigl( {{{\bar n}_g}(\eta)} - {1\over g} \Bigr) +
{\int_{-\beta\mu}^0} {d\eta} {1\over g}.\cr}
 \eqn\intcvar$$
Since the integrand of the second integral vanishes exponentially
at the lower limit of integration, this limit may be taken to
$-\infty$ at low temperature. This may then be combined with the
first integral while the third integral may be done, yielding
$$ {{\beta}^{p+1}} {\int_0^\infty}{d\epsilon} {{\epsilon}^p}
\,{{{\bar n}_g}(\epsilon-\mu)} = {1\over g}\,{1\over {p+1}}\,
(\beta\mu)^{p+1} + {\int_0^\infty}
{d\eta}\biggl({(\eta+\beta\mu)^p} {{{\bar n}_g}(\eta)} +
{(\beta\mu-\eta)^p}
\Bigl({{{\bar n}_g}(-\eta)} - {1\over g}\Bigr) \biggr)\eqn\intcomb$$
Expanding in powers of $\beta\mu$ using the binomial theorem,
we obtain
$$ {{\beta}^{p+1}} {\int_0^\infty}{d\epsilon} {{\epsilon}^p}
\,{{{\bar n}_g}(\epsilon-\mu)} =  {1\over g}\,{1\over {p+1}}\,
(\beta\mu)^{p+1} + {\sum_j}{(\beta\mu)^{p-j}}{p\choose j} {C_j}
\eqn\intexp$$
where we have isolated the pure numbers
$$ {C_j} =  {\int_0^\infty}
{d\eta} {{\eta}^j} \biggl( {{\bar n}_g}(\eta) +
{(-1)^j}\Bigl({{{\bar n}_g}(-\eta)} - {1\over g}\Bigr) \biggr)~.
\eqn\defints$$
Finally, then, we obtain
$$ {{\mu}_0^{D\over n}} =  {{\mu}^{D\over n}}
\biggl(1 + g(D/n)\,{\sum_j}{\biggl({T\over\mu}\biggr)^{j+1}}
{{-1+D/n}\choose j}{C_j} \biggr)
\eqn\cpexp$$
$$ E =  {E_0}      { \biggl({\mu\over{\mu_0}}\biggr) ^ {{D\over n}+1} }
\biggl(1 + g(1+D/n)\,{\sum_j}{\biggl({T\over\mu}\biggr)^{j+1}}
{{D/n}\choose j}{C_j} \biggr)
\eqn\enexp$$
where ${\mu_0}$ and ${E_0}$ are the zero-temperature chemical
potential and energy, respectively. Thus to second order in $T$
$$ {\mu} = {\mu_0}\Bigl(1 + g{C_0}\Bigl({T\over{\mu_0}}\Bigr)
- g(-1+D/n){C_1}{\Bigl({T\over{\mu_0}}\Bigr)^2}\Bigr)\eqn\cpexpt$$
$$E = {E_0}\biggl(1 + \Bigl(g(1+D/n){C_1} + {g^2}{C_0^2}
\Bigl({(D/n)(1+D/n)\over 2}-{(1+D/n)^2}\Bigr)\Bigr)
{\Bigl({T\over{\mu_0}}\Bigr)^2}\biggr)\eqn\enexpt$$
In the special case $g=1/2$, ${C_0}$
can be evaluated analytically.  Using the distribution
function \semiondist, one finds rather surprisingly
$$\eqalign{
({C_0})_{g=1/2} &= \int^\infty_0 du {1\over (1+ 4e^u )^{1\over 2}}
{}~+~ \int^0_{-\infty} du \bigl( {1\over (1+ 4e^u )^{1\over 2}} - 1 \bigr)\cr
&= {1\over 2} \int^\infty_2 {dv\over v } {1\over (1+v^2 )^{1\over 2} }
{}~+~ {1\over 2}
\int^2_0 {dv\over v } \bigl( {1\over (1+v^2 )^{1\over 2}} -1 \bigr) \cr
&= 0 \cr }
\eqn\semionshift$$
Hence, the $g=1/2$ specific heat to order $T$ is simply
$${\Bigl({C\over T}\Bigr)_{g=1/2}} = {(1+D/n)}
\,{{E_0}{C_1}\over \mu_0^2}\eqn\semspheat$$
${C_1}$ may be evaluated numerically: ${C_1}=1.6449$.
In the case of fermions ${C_n}=0$ for all even $n$
due to particle-hole symmetry, but
this is not the case for $g=1/2$: specifically, ${C_2}=1.2021$.
Remarkably, we find numerically that ${C_0}=0$ for
$g=1/3, 1/4$ as well. These results -- together with a duality
property to be demonstrated later which implies that
${C_0}=0$ for $g=2,3,4$ as well -- lead us to conjecture that
${C_0}=0$ for arbitrary $g$.

If the particle number is not conserved one has, immediately,
$$E = {\sum_i}{\epsilon_i}{n_i} = V {\int}{{{d^D}p}\over(2\pi)^D}
\,{\epsilon(p)}{{{\bar n}_g}(\epsilon(p))}
= \alpha V\, {T^{{D\over n}+1}}\, {\int}{d\eta}\,{{\eta}^{D\over n}}
\,{{{\bar n}_g}(\eta)}
\eqn\enernoncons$$
For ${1\over 2}$-ons with $D/n~=~1$,
the integral in \enernoncons\ takes the value 0.9870.

\chapter{Fluctuations, and a Perspective on the Assumptions}

Let us attempt to find the probabilities
for various occupation numbers of a single state,
for $g=1/2$.
Defining $f(n)e^{-n\beta (\epsilon - \mu )}$ as the probability for
$n$-fold occupancy, we derive from \semiondist\ the formal relation
$$
{\Sigma n f(n) e^{-n\beta (\epsilon - \mu )}
\over \Sigma f(n) e^{-n\beta (\epsilon - \mu )} }
{}~=~
{1\over ({1\over 4} + e^{2\beta (\epsilon - \mu )})^{1\over 2} }~.
\eqn\pathdis
$$
Matching coefficients and normalizing $f(0)=1$ we find
$f(1)=1$, $f(2)=1/2$, $f(3)=1/8$, $f(4)=0$, $f(5)=-1/128$, ... .
Clearly something has gone awry here!

The mathematical problem is as follows.
The dimension of the many-particle Hilbert space
$$W = {{(k+(1-g)(N-1))!} \over {N! ((k-1)-g(N-1))!}}\eqn\entcountred$$
for $N$ particles when the cell includes $k$ states
vanishes at $g={m\over{N-1}}$ where $m=k,k+1,\ldots ,k+N-1$.
This gives the correct result for fermions, namely
$W=0$ if $N>k$ since $g=1$ is then one of the zeroes of $W$.
However, $W$ does not vanish for general $g$ when $gN>k$. Indeed,
$W$ can be negative when $gN>k$. (To see this, consider the simple
case of $g=1/2$ for $N>2k$ and $N$ an even number.)
To get a sensible result for
the Hilbert space dimension, $W$, we must stipulate that
$W=0$ if $gN>k$, which complicates the minimization
of the entropy.
If we take a large cell size $k$
we can safely ignore this complication,
since $W$ is small for $gN>k$. For instance, if $g=1/2$,
$N=2k+2$, $W\sim -{1\over k}$ and if $N\gg 2k$, then
$W\sim -{1\over N}$.

Thus for $g$-ons
it is important to keep the cell size large, in contrast
to
the Fermi and Bose cases where one could take $k=1$
with impunity. In the fermion case, for instance, one has
$${p_0} + {p_1} =1\eqn\probtot$$
$${p_1} = {\overline n} = {\overline{n^2}} = \ldots = {\overline{n^l}}
\eqn\probmom$$
where ${p_i}$ is the probability of having
occupation number $i$.  One can calculate moments of the occupation
number, $\overline {n^m}$, by differentiating the distribution function.
The closure property, ${\overline{n^l}} = {\overline{n^m}}$ holds because
one can take $k=1$ and the only allowed occupation numbers are
$0$ and $1$. For ${1\over 2}$-ons, one must keep $2k$ equations
for ${\overline n}, {\overline{n^2}},\ldots,{\overline{n^{2k}}}$
in terms of ${p_1},{p_2},
\ldots,{p_{2k}}$. The closure equation which expresses
${\overline{n^{2k+2}}}$ in terms of the first $2k$ moments is only
satisfied to $O({1\over k})$.

One must therefore exercise some care in using the distribution
functions derived from \impldistfcn. The cell size must not
become too small in view of the preceding
paragraph, but it must also not become too
large, because if the energy spread within one cell
becomes comparable to the temperature then the notion
of a characteristic energy for the cell becomes invalid.
It is amusing that negative probabilities appear in this problem in
a natural and meaningful way: it is necessary that negative probabilities
appear in the description of
small cells, if independent addition of many such is to generate
the correct average occupancy for large cells!

Our mathematical problem reflects a fundamental implicit
{\it physical\/} assumption
in the derivation of $g$-on statistical mechanics.
For bosons or fermions the fundamental assumption of symmetry or
antisymmetry of the
wave-function holds rigorously and locally
in momentum space.
This is enough to allow
one to derive the appropriate statistical
distribution for an ideal gas locally in phase space.
To derive $g$-on statistical mechanics as above one must
assume that the generalized exclusion principle operates on states
of nearby energy, and as we have seen one must also
take a cell size not too small.  Without attempting a rigorous discussion,
we can identify qualitatively the
physical circumstance under which
these assumptions
become plausible.  It is that the effective interaction which reduces
the Hilbert space dimensions should be
essentially local in momentum space.  Then one may apply the counting
arguments to cells containing all the states in a small range of
momenta: this will be
a number of states proportional to the volume, all with
essentially the same energy.

The known examples of $g$-ons have this character.
Ideal $g$-on statistical mechanics
operates in the Calogero-Sutherland models [\bernard ],
which feature
long-range interactions.
Anyon models (that is, $2+1$-dimensional systems
with a Chern-Simons gauge field) feature
interactions which are singular for nearby momenta,
resulting in a shift of the
allowed values of the relative angular momentum between
two particles, $l \rightarrow l+\alpha$.
Since two particles occupying the
same state must have vanishing relative angular momentum,
an anyon excludes its state from further occupation.
This is certainly
local exclusion, but the second condition is not satisfied:
an anyon in any other state also has its relative angular momentum
shifted. Hence anyons are not
ideal $g$-ons, but {\it interacting\/} $g$-ons.  Whether the ideal
$g$-on statistical mechanics provides a useful first approximation
in this case, is a question needing further investigation.  Hard-core
bosons on a lattice have $g=1$ according to Haldane's definition, but
are far from being 1-ons according to our definition,
and their behavior is
poorly approximated by fermions:
one expects them to
Bose condense rather than to form a Fermi surface at low temperatures.

%%%%%%%%%%%%%%%%%%%%%%%%

To conclude this discussion, let us
finally display the first-order fluctuations concretely, for $g=1/2$.
Using the standard identity
relating ${\overline {(\Delta n)^2}}$ to the derivative of ${\bar n}$
with respect to the chemical potential, we find:
$${\overline {(\Delta n)^2}} = {\bar n} (1 - {1\over 4}{{\bar n}^2})
\eqn\fluct$$
Thus we find that ${1\over 2}$-ons
have sub-Poissonian statistics,
as do fermions ${\overline{(\Delta n)^2}} = {\bar n}(1-{\bar n})$.
In contrast, bosons are super-Poissonian, ${\overline{(\Delta n)^2}} =
{\bar n}(1+{\bar n})$.

\chapter{Duality}

We alluded earlier to a duality property that relates statistics
$g$ and $1/g$. This property may be seen by rewriting the implicit
relation for the distribution function at statistics $g$,
$${({{\bar n}_i})_g}\,{e^{\beta({\epsilon_i}-\mu)}} =
{\Bigl(1-g{({{\bar n}_i})_g}\Biggr)^g}
{\Bigl(1+(1-g){({{\bar n}_i})_g}\Bigr)^{1-g}}
\eqn\impl$$
in the form,
$${\Bigl(1-g{({{\bar n}_i})_g}\Bigr)}
\,{e^{-{{\beta}\over g}({\epsilon_i}-\mu)}} =
{({{\bar n}_i})_g^{1/g}}
{\Bigl(1+(1-g){({{\bar n}_i})_g}\Bigr)^{1-{1\over g}}}
\eqn\implm$$
or,
$${\Bigl(1-g{({{\bar n}_i})_g}\Bigr)}
\,{e^{-{{\beta}\over g}({\epsilon_i}-\mu)}} =
{\biggl({1\over g} - {1\over g}
\Bigl(1 - g({{\bar n}_i})_g\Bigr)\biggr)^{1/g}}
{\biggl({1\over g}+\Bigl(1-{1\over g}\Bigr)
{\Bigl(1 - g({{\bar n}_i})_g\Bigr)}\biggr)^{1-{1\over g}}}
\eqn\implmm$$
But the implicit relation for the distribution function
at statistics $1/g$ reads
$${({{\bar n}_i})_{1/g}}\,{e^{\beta({\epsilon_i}-\mu)}} =
{\biggl(1-{1\over g}{({{\bar n}_i})_g}\biggr)^{1/g}}
{\biggl(1+\Bigl(1-{1\over g}\Bigr){({{\bar n}_i})_g}\biggr)^{1-{1\over g}}}
\eqn\implginv$$
Hence,
$$1 - g{{\bar n}_g}(\beta(\epsilon-\mu))
= {1\over g}{{\bar n}_{1/g}}(-{\beta\over g}(\epsilon-\mu))
\eqn\petunia$$
This duality relates the distribution of holes in the $g$-on
distribution (where full filling is at $\overline n_g = 1/g$) to
that of $1/g$-ons at $g$ times the temperature, or alternatively
$1/g$ times the energy and chemical potential.

This duality is reminiscent of the one found in Chern-Simons models.
There one describes anyons as charge $Q=q$ objects which acquire
a proportional flux
$\Phi = q/\mu$, where $\mu$ is the Chern-Simons coupling, and thereby have
their statistics transmuted by $\Delta {\theta \over \pi} =
Q\Phi = q^2/\mu$.
The fundamental flux tubes then have flux $\Phi = 1/q$, charge
$Q=\mu /q^2$, and the inverse statistics.
The thermal duality \petunia\  is also reminiscent of the
$g\rightarrow 1/g$ duality in the Calogero-Sutherland models [\csrefs ].
Indeed, {\it for these models\/} thermal duality as discussed here
follows from the known
coupling-constant duality.  The concordance of the general thermal
duality for abstract $g$-on statistics with the more specific and
complete duality
for Calogero-Sutherland quanta, vividly confirms the idea that these models
embody ideal $g$-on statistics.

\chapter{Remarks on the Mott Problem}

The application of $g$-on statistics to one dimensional systems, even
for the soluble models where it is formally correct, is
not straightforward.  Indeed in one space
dimension the application of Fermi
statistics to derive the low-energy properties of systems of fermionic
quasiparticles has to be carefully considered, since interactions can
change the properties qualitatively.  The Fermi liquid must be considered
as one special case of the generic Luttinger liquid [\luttinger ].  Thus
for example the thermodynamic properties of edge excitations in the
fractional quantized Hall states are not correctly reproduced
by the ideal
$g$-on formulas, even though the electrons are (for example)
formally $m$-ons in the $\nu = 1/m$ state.
Of course the bulk filling fraction is nicely consistent with $1/g$
filling of the magnetic band, but this is a much weaker statement.

More speculative, but if correct probably
more useful, is the possibility of applications
to systems in higher dimensions.  For in higher dimensions
the phase space arguments of Landau apply, as in his justification
of Fermi liquid theory, and make it plausible that (unlike in one
space dimension) the approximation of non-interacting quasiparticles
is accurate at low temperature.

There is a class of insulating materials, the Mott insulators, which
are anomalous from the point of view of band or Fermi-liquid theory.
They are insulators when their valence band is precisely half filled.
 From the point of view of this paper, it is natural to hypothesize
that in
these materials the electrons are behaving as 2-ons
[\slawek ].  The most important
qualitative feature of Mott insulators, that is the
existence of a gap at exactly half filling, follows directly.
Such behavior
is suggested, but certainly {\it not\/}
proved,
by the idea (formalized in the t-J model) that because of
strong on-site repulsion a single electron excludes two states -- namely
states of both spins -- from its lattice site.
As we have taken pains to emphasize, what is needed is local repulsion in
momentum space, which could arise directly from a long-range force or
indirectly through correlation effects.

In any case, our hypothesis leads
to the statistical-mechanical consequences derived earlier, which could be
tested in experiments or by numerical work on models.
Most interesting are effects which arise just {\it below\/} half filling.
The central consequence is the
existence at $T=0$ of a ``Fermi surface'' of anomalous size, and with
anomalous values of the specific heat, Pauli susceptibility, ... .
There {\it are\/} quantitative anomalies in the experimentally observed
normal state specific heat of the CuO-based
superconductors [\ginz ],
and in the size of the ``Fermi surface'' in the t-J model
as calculated using high-temperature expansions [\highT ] and
variationally [\var ].  (In these calculations, the nominal Fermi surface
is identified from a strong feature in the density-density correlation
function.)  These anomalies are at least roughly consistent
with the 2-on hypothesis: the specific heat is substantially larger, and
the volume of the ``Fermi surface'' is roughly twice as large, as
would be expected
for ordinary fermions.

\endpage

\refout

\endpage

\end